\documentclass[pdftex,twocolumn,epjc3]{svjour3}
\usepackage{graphics}
\usepackage{times,mathptm}
\usepackage{amsmath,amsfonts,amssymb, amsbsy}
\usepackage{graphicx,graphics,epsfig}
\usepackage{fancyhdr,fancybox}
\usepackage{verbatim}
\usepackage{xcolor}
\usepackage{multirow}
\usepackage{enumerate}
\usepackage{cite}
\usepackage{float}
\usepackage{array}
\usepackage[colorlinks]{hyperref}
\usepackage{cuted}
\usepackage{soul}
\usepackage{bm}

\newcommand{\vv}{\mathbf{v}}
\newcommand{\vp}{\mathbf{p}}
\newcommand{\vh}{\mathbf{h}}
\newcommand{\vu}{\mathbf{u}}
\newcommand{\ve}{\mathbf{e}}
\newcommand{\vr}{\mathbf{r}}

\def\Gs{\sigma}

\renewcommand{\d}{\mathop{}\mathopen{}\mathrm{d}}
\newcommand{\D}{\mathop{}\mathopen{}\mathrm{D}}

\DeclareMathOperator{\sign}{sign}
\newcommand*\diff{\mathop{}\!\mathrm{d}}

\journalname{}
\def\makeheadbox{\relax}

\begin{document}
\title{Spontaneous flow created by active topological defects}
\author{Louis Br\'ezin\thanksref{addr1,addr2}
        \and
        Thomas Risler\thanksref{e2,addr1}
        \and
        Jean-Fran\c cois Joanny\thanksref{e1,addr1,addr2}
}

\thankstext{e1}{e-mail: jean-francois.joanny@college-de-france.fr}
\thankstext{e2}{e-mail: thomas.risler@curie.fr}

\institute{Institut Curie, Universit\'e PSL, Sorbonne Universit\'e, CNRS UMR168, Laboratoire Physico Chimie Curie, 75005 Paris, France\label{addr1}
          \and
          Coll\`ege de France, 75005 Paris, France\label{addr2}
}

\date{February 1, 2022}

\abstractdc{Topological defects are at the root of the large-scale organization of liquid crystals. In two-dimensional active nematics, two classes of topological defects of charges $\pm1/2$ are known to play a major role due to active stresses. Despite this importance, few analytical results have been obtained on the flow-field and active-stress patterns around active topological defects. Using the generic hydrodynamic theory of active systems, we investigate the flow and stress patterns around these topological defects in unbounded, two-dimensional active nematics. Under generic assumptions, we derive analytically the spontaneous velocity and stall force of self-advected defects in the presence of both shear and rotational viscosities. Applying our formalism to the dynamics of monolayers of elongated cells at confluence, we show that the non-conservation of cell number generically increases the self-advection velocity and could provide an explanation for their observed role in cellular extrusion and multilayering. We finally investigate numerically the influence of the Ericksen stress. Our work paves the way to a generic study of the role of topological defects in active nematics, and in particular in monolayers of elongated cells.}
\maketitle

Topological defects are singularities in the topology of the order-parameter field in a phase of broken continuous symmetry~\cite{MerminTopological1979,kleman1983}.
They were first observed in nematic liquid crystals as early as 1904 by Lehman~\cite{lehmann1904} and sparked theoretical interest later by Frank and Kleman for their role in the mechanical and optical properties of liquid
crystals~\cite{kleman1983,kleman1989}. In two-dimensional nematic liquids, topological defects are point-like singularities of the director-field orientation. They are classified by the defect strength or topological charge $S$, which is equal to the number of full rotations of the liquid-crystal director on a closed path of total angle $2\pi$ around the defect core~\cite{OswaldNematic2005,HarthTopological2020}. Due to the nematic symmetry, this topological charge is either an integer or a half-integer. The regions that surround a topological defect present strong gradients of the director-field orientation, with high energetic costs. Because of their relatively lower energy, the defects that are mostly observed in passive nematic liquid crystals have topological charge $S=\pm1/2$, which is the lowest possible charge in absolute value.
    
Largely driven by biophysical applications, much attention has been attracted more recently to active matter~\cite{JulicherActive2007,marchetti2013}. In active nematics, numerical simulations~\cite{giomi2013,thampi2013} as well as analytical analyses~\cite{pismen2013} have revealed a very rich dynamics, dominated by the creation, motion, and annihilation of defects. Topological defects in living systems have been first observed in 1968 by Elsdale in monolayers of fibroblast cells~\cite{elsdale1968} and, since then, the topological charge of the defects has helped to identify the nematic or polar nature of cellular tissues~\cite{kemkemer2000,duclos2017a,saw2018,blanch-mercader2020,blanch-mercader2020a}. Recent works looked at the influence of topological defects on the onset of three-dimensional morphogenesis~\cite{maroudas-sacks2021,metselaar2019}. Some material properties can also be extracted from the observation of the orientation of the director field around topological defects~\cite{blanch-mercader2020,blanch-mercader2020a}.

The properties of topological defects in active nematic liquid crystals are very different from the properties of their equilibrium counterparts. The active stress in an active nematic is proportional to the nematic orientational tensor. As regions surrounding a topological defect have pronounced nematic orientation gradients, they correspond to regions of strong active force density, which induce local flows around the defect. The structure of this flow depends on the topological charge of the defect, but defects in general are associated with vortices of the flow field~\cite{giomi2013,thampi2013}. If the director field around the defect breaks a left-right symmetry along a particular direction, and in the absence of any external force, the flow created by active stress gradients drags the defect along that direction with a finite velocity, limited by viscous dissipation. Defects of topological charge $S=-1/2$ have a three-fold symmetry. Consequently, they have no preferred direction of motion and passively diffuse.  On the contrary, defects of topological charge $S=+1/2$ break a `head-tail' symmetry. In active nematics, they generically move along their head-tail axis, with a direction dictated by their contractile (motion toward the tail) or extensile (motion toward the head) nature.

Experimental observations on nematic cell monolayers show that topological defects are preferential sites of cell extrusion~\cite{saw2017,kawaguchi2017} and multilayering~\cite{sarkar2021}. Similar observations have been made in bacterial monolayers~\cite{copenhagen2021}. In a recent work by T.~Sarkar and colleagues~\cite{sarkar2021} on the spontaneous organization of cell monolayers on a solid substrate, the authors measured the flow patterns around topological defects of charges $\pm 1/2$. Some of the $+1/2$ topological defects were found to be pinned at specific locations on the substrate, while others had a spontaneous motion as expected in active nematics.

In the current paper, we study theoretically the flows generated by active, two-dimensional topological defects of charges $S=\pm 1/2$. One key ingredient of the hydrodynamics of nematic liquid crystals is the backflow, the hydrodynamic motion induced by the rotation of the director field~\cite{degennes-prost}. Analytical results on topological defects have been obtained by neglecting this backflow~\cite{giomi2014,Ronning2022}, while numerical studies have considered the coupling between the flow and the director fields in passive~\cite{toth2002} as well as active~\cite{giomi2013,giomi2014} nematic defects. We provide here analytical solutions that account explicitly for these effects. We compute the force necessary to stall an active topological defect of charge $S=+1/2$ and investigate the link between the two-dimensional 
flow patterns and the trigger of multilayering. We limit our study to a single, isolated defect in an infinite plane. Thereby, we rely on analytical solutions of the active hydrodynamic equations that depend on the bulk properties of the cellular tissue only. Performing a perturbation analysis around the passive defect orientation allows us to obtain analytical solutions for weakly active systems. We also estimate the size of the spatial domain over which this hypothesis is valid.

The paper is organized as follows: section~\ref{sec_governing_eq} introduces the hydrodynamic equations of active nematics by giving the constitutive equations and computing the active force density generated by a topological defect of charge $S=\pm 1/2$. We then study in section~\ref{sec_flow_simple} the dynamics of the defect, assuming that the nematic orientation is unperturbed compared to that of a passive defect of the same charge. We focus on $+1/2$ defects, but computations for $-1/2$ defects are detailed in~\ref{app_m12}. We compute the flow field and the self-advection velocity of the defect in section~\ref{sec_self_advection_simple}. In section~\ref{sec_pinning_simple}, we compute the force necessary to pin a topological defect to a specific location on the substrate. In section~\ref{sec_division}, we study the effect of inhomogeneous cell divisions and cell deaths or extrusions on the self-advection velocity of the defect, in the case where the rate of cell production is linearly coupled to the differential pressure field. In section~\ref{sec_rotational_dyn}, we address the role of the rotational viscosity, which induces feedback of the flow on the nematic orientation. We first ignore the elastic Ericksen stress and set the flow-alignment parameter to zero in section~\ref{sec_self_advection_gamma}. This allows for the computation of the flow and the self-advection velocity to linear order in activity analytically. Within the same hypothesis, the crucial role of the backflow on the stall force is detailed in section~\ref{sec_pinning_gamma}. The first-order correction to the director orientation as well as the spatial domain of validity of our calculations are discussed in section~\ref{sec_orientation}. Finally, we investigate numerically the role of the Ericksen stress on the self-advection velocity in section~\ref{sec_ericksen}.

\section{Hydrodynamic equations}\label{sec_governing_eq}

We consider an active nematic deep in the nematic phase where the order parameter has a constant modulus. To describe its dynamics, we make use of the hydrodynamic theory of active nematic gels~\cite{kruse2005,joanny2007,marchetti2013}. This very general framework is based on symmetries and conservation laws~\cite{joanny2009} and has been shown to adequately describe monolayers of elongated cells that self-organize into a nematic phase~\cite{duclos2014,duclos2018b,saw2018,alert2020}. More specifically, we aim at describing the dynamics of nematic topological defects as observed experimentally in monolayers of myoblast cells~\cite{sarkar2021}. Because we are interested in monolayers of cells at confluence, we consider the active nematic as a one-constituent dense phase, with a fixed cell density.

Within the active hydrodynamic framework, the monolayer is described by a velocity field $\vv$ and a director field $\vp$. The two-dimensional Frank-Oseen free energy $F$ associated to gradients of the director-field orientation reads, in the one-constant approximation where the splay and bend constants are equal~\cite{degennes-prost}:
\begin{equation}\label{eq_Frank_energy}
F = \int{\d x \: \d y \left[ \frac{K}{2} \textrm{Tr}\left[\left(\nabla\otimes\vp\right)\cdot\left(\nabla\otimes\vp\right)^{\rm T}\right] - \frac{1}{2}h_\parallel^0\, \vp^2 \right]}\, ,
\end{equation}
where $\otimes$ denotes the tensorial product, and $h_\parallel^0$ is a Lagrange multiplier to ensure that the director is a unit vector.

The evolution of the velocity and director fields is described by two coupled vectorial equations. The first equation is the force-balance equation, which reads, at steady state:
\begin{equation}\label{eq_FB_formal}
\nabla \cdot \bm{\Gs}=\mathbf{0}\, .
\end{equation}
Inertia has been neglected since the Reynolds number for typical cellular systems is much smaller than one.

The total stress tensor is further split into a passive $\bm{\Gs}^{\rm p}$ and an active $\bm{\Gs}^{\rm a}$ contribution. The passive contribution is given by the hydrodynamic theory of passive nematics~\cite{degennes-prost,martin1972,forster1971} as:
\begin{align}\label{eq_stress_passive}
\bm{\Gs}^{\rm p} =& 2\eta\,\tilde\vu + \frac{\nu}{2}\left(\vh\otimes\vp + \vp\otimes\vh - (\vp\cdot\vh)\,\mathbf{1}\right) \nonumber \\
    &+ \frac{1}{2}\left(\vh\otimes\vp - \vp\otimes\vh\right) + \tilde{\bm{\Gs}}^{\rm E} - P\,\mathbf{1}\, .
\end{align}
Here, $\eta$ is the shear viscosity; $\tilde\vu$ is the traceless part of the symmetric part of the velocity-gradient tensor $\vu=(\nabla\otimes\vv+(\nabla\otimes\vv)^{\rm T})/2$; $\nu$ is the so-called flow-alignment parameter; $\vh=-\delta F/\delta\vp$ is the orientational field; $\tilde{\bm{\Gs}}^{\rm E}$ is the traceless part of the Ericksen stress tensor $\bm{\Gs}^{\rm E}$, which generalizes thermodynamic pressure for anisotropic systems~\cite{degennes-prost,chaikin1995}; and $P$ is the pressure, which is a Lagrange multiplier that ensures the incompressibility condition $\nabla \cdot \vv = 0$, with $\mathbf{1}$ the identity tensor. The elastic force density associated with the Ericksen stress is obtained from the Gibbs-Duhem relation~\cite{julicher2018,joanny2007,degennes-prost}
\begin{equation}\label{eq_Gibbs_Duhem}
\nabla \cdot \bm{\Gs}^{\rm E} =-(\nabla\otimes\vp)\cdot\vh\, .
\end{equation}

The active contribution is linked to the existence of a local cell alignment along the direction of the director field $\vp$. It reads
\begin{equation} \label{eq_stess_active}
\bm{\Gs}^{\rm a} = -\zeta \Delta \mu\left(\vp\otimes\vp - \frac{\mathbf{1}}{2}\right)\, ,
\end{equation}
where $\zeta \Delta \mu$ is a scalar quantity that measures the activity of the system. It is the product of a difference in chemical potential $\Delta\mu$, which is positive, and an Onsager coefficient $\zeta$, which can have either a positive or a negative value. A positive value of $\zeta$ corresponds to an extensile active stress, in which the cells push along their long axis. The stress $\zeta \Delta \mu$ is negative in a contractile system where the cells pull along their long axis.

The second equation describes the evolution of the director field $\vp$:
\begin{equation}\label{eq_dyn_p}
    \frac{\D \vp}{\D t} = \frac{1}{\gamma}\vh - \nu\,\vu\cdot\vp\, .
\end{equation}
Here, $\D \vp/\D t = \partial_t \vp + (\vv\cdot\nabla)\vp + \bm{\omega}\cdot\vp$ is the co-moving co-rotational derivative of the director, where $\bm{\omega}=(\nabla\otimes\vv-(\nabla\otimes\vv)^{\rm T})/2$ is the vorticity tensor, and $\gamma$ is the rotational viscosity. The orientational field $\vh= -\delta F/ \delta \vp$ is associated to changes of the free energy with respect to the director $\vp$. It is parallel to $\vp$ in a non-flowing steady state. It is convenient to introduce the components of the orientational field, parallel and perpendicular to $\vp$, $h_\parallel$ and $h_\perp$, respectively. The component $h_\perp$ controls the orientation of the director $\vp$, whereas the component $h_\parallel$ controls the modulus of $\vp$~\cite{kruse2005}.

\begin{figure}
    \centering
    \includegraphics[width=0.45\textwidth]{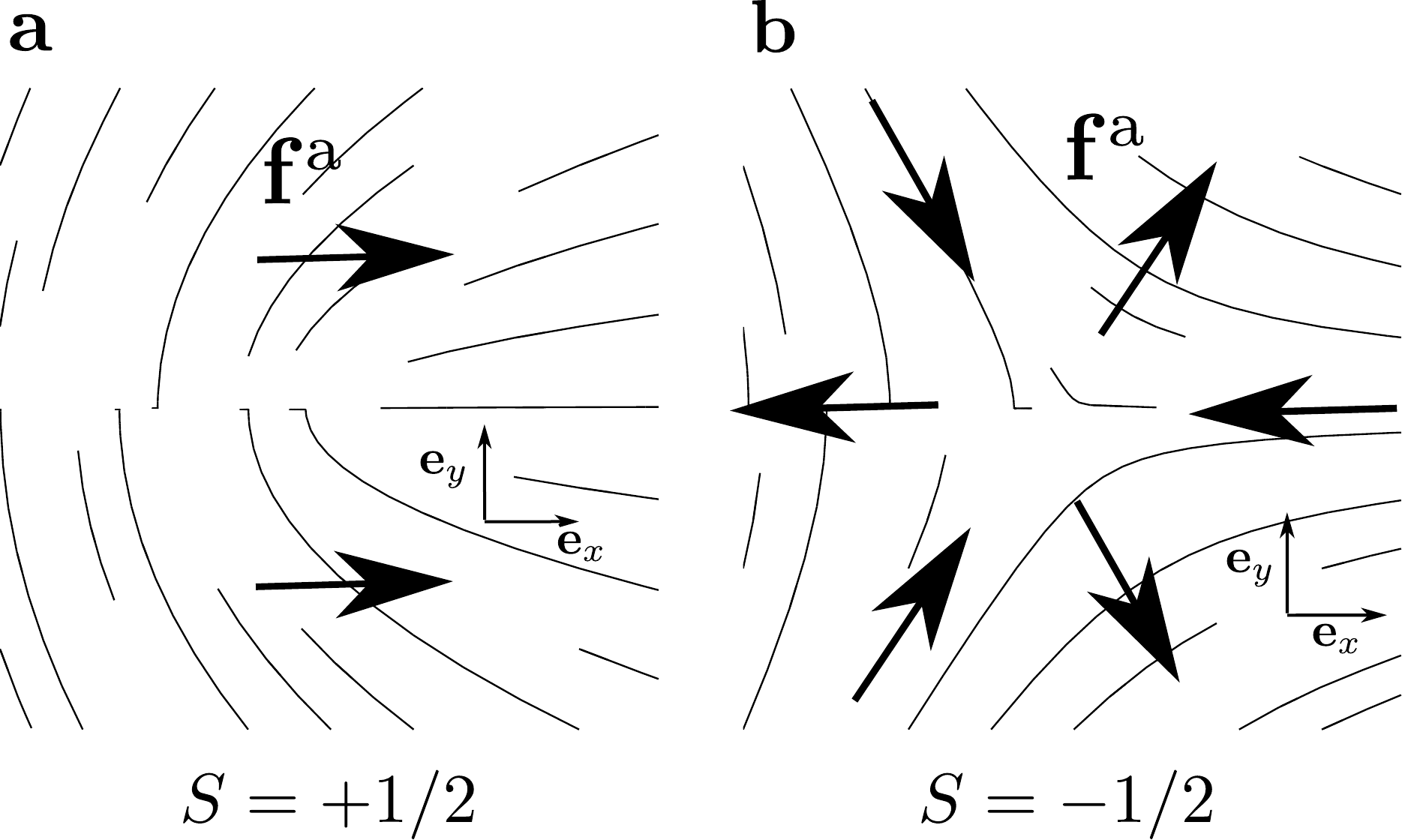}
    \caption{Passive orientation of the director field (solid lines) and resulting active force density (arrows) in the contractile case $\zeta \Delta
    \mu < 0$, for $+1/2$ (\textbf{a}) and $-1/2$ (\textbf{b}) topological defects. The origin of the polar angle $\theta$ is set by $\mathbf{e}_x$.}
    \label{fig_forces}
\end{figure}

In the following, we define the orientation angle $\varphi$ of the director $\vp$ as $\vp = \cos{\varphi}\,\ve_x + \sin{\varphi}\,\ve_y$ in the fixed reference frame $\{\ve_x,\ve_y\}$. For passive nematics, the equilibrium configurations correspond to minima of the Frank-Oseen free energy given in eq.~\eqref{eq_Frank_energy}, which leads to $\Delta \varphi = 0$. This equation has the solution $\varphi(r,\theta) = S\,\theta$ for a topological defect of charge $S$, using polar coordinates $(r,\theta)$ centered at the defect singularity. Defects with the lowest absolute topological charge $S=\pm 1/2$ are the most stable. They are shown in fig.~\ref{fig_forces}.
In free monolayers of nematic tissue, they are the only observed defects.

For a director field corresponding to a passive defect, using the polar coordinates $(r,\theta)$ in the Cartesian basis of vectors $\{\ve_x,\ve_y\}$, the active force density reads:
\begin{equation}
    \begin{aligned}
    \mathbf{f}^a = \nabla\cdot\bm{\sigma}^{\rm a} = -\frac{\zeta \Delta \mu}{2r}
    \begin{cases}
        \ve_x\ , & S=+1/2\\
        -\cos{2\theta}\ \ve_x + \sin 2\theta\ \ve_y\ , &
        S=-1/2
    \end{cases}
    \end{aligned}
    \label{eq_active_force}
\end{equation}
As represented in fig.~\ref{fig_forces}, the active force density for a $+1/2$ topological defect is along the axis of symmetry. This drives a self-advection of the defect `from head to tail' in the contractile case and `from tail to head' in the extensile case. For $-1/2$ defects, because of the three-fold symmetry of the director field, the net total force on a disk of radius $R$ vanishes, such that the default, even tough active, can only diffuse.

\section{Topological defects in the limit of vanishing rotational viscosity}\label{sec_flow_simple}

Throughout this section, we consider the case of vanishing rotational viscosity. In this case, the molecular field vanishes to satisfy eq.~\eqref{eq_dyn_p}, so that $\delta F/\delta\vp=\mathbf{0}$ and the director orientation satisfies the equilibrium condition for passive nematics $\Delta \varphi = 0$, illustrated in fig.~\ref{fig_forces}. The force density associated to the Ericksen stress as given by eq.~\eqref{eq_Gibbs_Duhem} vanishes. The Ericksen stress can then be ignored and the passive stress as given by eq.~\eqref{eq_stress_passive} reduces to ${\bm \sigma}^p =2\eta\,\tilde\vu - P\,\mathbf{1}$.

\subsection{Flow field and self-advection velocity of a $+1/2$ defect}\label{sec_self_advection_simple}

Under this assumption, we now compute the velocity field of an isolated $+1/2$ topological defect in an infinite domain. We write the equations of motion in the reference frame of the defect. Force balance reads
\begin{equation} \label{eq_FB_vector}
\eta \Delta \mathbf{v} - \mathbf{\nabla} P - \xi (\mathbf{v}+\mathbf{v_0})  - \frac{\zeta \Delta\mu}{2r}\ve_x = 0\ ,
\end{equation}
where $\xi$ is the friction coefficient of the cell layer with the underlying substrate and $\vv_0$ the self-advection velocity of the defect with respect to the substrate. This bulk equation is supplemented by two boundary conditions on the velocity field: it vanishes at the core of the defect in the reference frame of the defect $\mathbf{v}(0,\theta)=\mathbf{0}$, and it vanishes at infinity in the reference frame of the substrate $\mathbf{v}(\infty,\theta)= - \mathbf{v}_0$.

Considering the incompressibility condition $\nabla \cdot \vv = 0$, eq.~\eqref{eq_FB_vector} is solved by introducing the stream function $\psi(r,\theta)$, defined by $\mathbf{v} = \mathbf{\nabla} \times (\psi\ \ve_z)$, where $\ve_z=\ve_x\times\ve_y$. We further adimensionalize the equations using the length scale $L = \sqrt{\eta/\xi}$ and the time scale $\tau = 2\eta/(|\zeta|\Delta\mu)$. In the following, all dimensionless quantities are denoted with a tilde. Taking the curl of eq.~\eqref{eq_FB_vector}, the dimensionless stream function $\tilde{\psi}=(\tau/L^2)\psi$ satisfies
\begin{equation}\label{eq_psi_4th_order}
\Delta \left[\Delta \tilde{\psi}  - \tilde{\psi}\right] = -s\frac{\sin \theta}{\tilde{r}^2}\ ,
\end{equation}
where $\tilde r=r/L$, $s=\sign(\zeta)$ and $\Delta$ denotes the Laplace operator with respect to the reduced variable $\tilde r$.

\begin{figure*}
    \centering
    \includegraphics[width=0.9\textwidth]{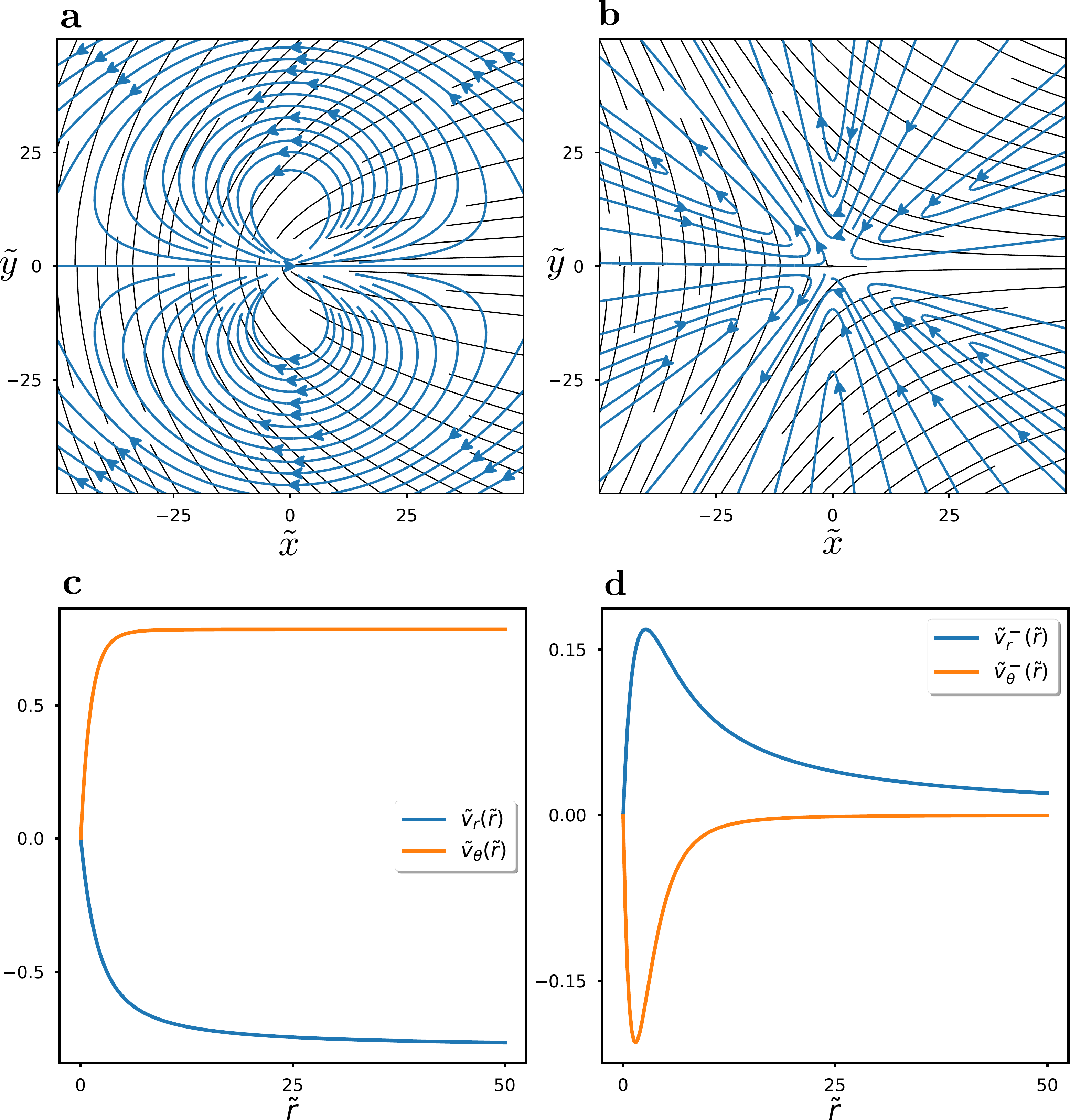}
    \caption{\textbf{a,b} Flow structure around active $+1/2$ (\textbf{a}) and $-1/2$ (\textbf{b}) topological defects. Thin, solid black lines represent the orientation of the director field. Thick, arrowed blue lines are the directed streamlines of the velocity field in the reference frame of the substrate for a contractile nematic. Vortices around $+1/2$ defects have a rotational direction that depends on the sign of $\zeta$. The three-fold symmetry for $-1/2$ defects does not create self-advection contrary to $+1/2$ defects. \textbf{c,d} Dimensionless azimuthal (orange) and radial (blue) velocities as a function of the dimensionless distance to the core $\tilde r$ in the reference frame of the defect. For $+1/2$ defects (\textbf{c}), this velocity vanishes at the core and tends to a finite limit at infinity. For $-1/2$ defects (\textbf{d}), this velocity vanishes both at the core and at infinity.}
    \label{fig_velocity_field}
\end{figure*}

A $+1/2$ topological defect is symmetric with respect to the $x$-axis. Therefore, the radial component of the velocity field is an even function of the polar angle $\theta$, and the stream function an odd function of that variable. Given that eq.~\eqref{eq_psi_4th_order} is linear in $\tilde{\psi}$ with a forcing term in $\sin{\theta}$, only the Fourier mode $n=1$ contributes non trivially to the solution, and the stream function can be written as $\tilde\psi(\tilde r,\theta) = \tilde\psi(\tilde r) \sin \theta$. The dimensionless velocity field has then polar components $\tilde v_r(\tilde r,\theta) = \tilde v_r(\tilde r) \cos \theta$ and $\tilde v_\theta(\tilde r,\theta) = \tilde v_\theta(\tilde r) \sin \theta$, which can be derived directly from the generic solution of eq.~\eqref{eq_psi_4th_order} for $\tilde{\psi}$, as detailed in~\ref{app_velocity_wo}. Imposing that the solution is non-divergent at infinity and vanishing in $\vr=\mathbf{0}$, the solution for $\tilde v_r$ and $\tilde v_\theta$ read:
\begin{align}\label{eq_v_final}
    \tilde{v}_r(\tilde r) =& -
    \frac{s}{\tilde r} \left\{I_1(\tilde r)
        \int_{\tilde r}^{+\infty}{K_1(u)(-\frac{\pi}{4}u^2 +
        u)\:
        \d u}   \right. \nonumber\\
         &\qquad  +\left. K_1(\tilde r)  
         \int_0^{\tilde r}{I_1(u)(-\frac{\pi}{4}u^2 + u)\: \d u} \right\}\\
    \tilde{v}_\theta(\tilde r) =&s \left\{\left( 
        I_0(\tilde r)-\frac{ I_1(\tilde r)}{\tilde r} \right)
     \int_{\tilde r}^{+\infty}{K_1(u)(-\frac{\pi}{4}u^2 + u)\: \d u}\right.\nonumber\\
         &\left. - \left(K_0(\tilde r) + \frac{K_1(\tilde r)}{\tilde r} \right)
         \int_0^{\tilde r}{I_1(u)(-\frac{\pi}{4}u^2 +
        u)\: \d u}\right\}  \ ,
\end{align}
where $I_1(u)$ and $K_1(u)$ are the modified Bessel functions of the first and second kind, respectively~\cite{abramowitz1964} (see~\ref{app_bessel}). The self-advection velocity can then be computed. Taking the limits of eq.~\eqref{eq_v_final} at infinity leads to
\begin{equation}\label{eq_vdefect}
\vv_0 = - \frac{\pi}{4} \times \frac{\zeta \Delta \mu}{2\sqrt{\xi\eta}}\ \ve_x\ ,
\end{equation}
in physical units. This result is similar to that of ref.~\cite{Ronning2022}. A representation of this velocity field as well as of its amplitude as a function of $\tilde r$ is shown in fig.~\ref{fig_velocity_field}a,c.
Corresponding expressions for $-1/2$ defects are given in~\ref{app_m12_velocity} and plotted in fig.~\ref{fig_velocity_field}b,d.
The solution~\eqref{eq_v_final} depends only on intrinsic properties of the system and can therefore be directly compared with experimental measurements to determine intrinsic parameters in experimental systems.

Because of hydrodynamic screening, the velocity field decays as $1/r$ at infinity in the substrate reference frame, as does the active-force density. The total active force diverges linearly with system size but is balanced by the total friction force. These divergences exist here because we consider a single defect in an unbounded space. For a pair of defects of opposite charges $+1/2$ and $-1/2$, for example, the active-force density 
and the velocity field decrease as $1/r^2$.
The pressure field is obtained from the force-balance eq.~\eqref{eq_FB_vector}:
\begin{equation}\label{eq_pressure}
P(r,\theta) = -\frac{\zeta \Delta \mu}{2} \cos \theta + P_0\ ,
\end{equation}
where $P_0$ is a reference pressure.
The active stress creates a differential pressure between the head and the tail of the defect. In the contractile case ($\zeta \Delta \mu<0$),
the pressure is higher at the tail of the defect, and it is the opposite in the extensile case.

\subsection{Stall force of a $+1/2$ defect}\label{sec_pinning_simple}

Sarkar \emph{et al.} observed two classes of $+1/2$ topological defects in confluent monolayers of C2C12 mouse myoblasts~\cite{sarkar2021}: motile topological defects with a flow characteristic of self-advected defects, as presented in sec.~\ref{sec_self_advection_simple}, and stalled, non-motile defects. These defects where not immobilized by jamming, as it can occur in other cellular systems~\cite{garcia2015}, as significant flows were observed around them. In addition, a one-to-one correspondence was observed between non-motile defects and the initiation of a second layer of cells due to extrusion of cells from the first layer, a process referred to as multilayering. Immotile $+1/2$ defects in an active monolayer can only exist in the presence of an external force that balances the total active force, other than the friction drag that structurally vanishes for a vanishing cell-velocity field.

We determine in this section the local external force applied on the defect core necessary to pin an active, $+1/2$ defect. We introduce a core region of the defect of finite size $a$ of the order of the correlation length of the nematic order~\cite{degennes-prost}, which we consider as a region where there is no nematic order. We therefore model the monolayer as divided into two regions: perfect nematic order with $\vp^2=1$ as in section~\ref{sec_self_advection_simple} for $r>a$, and isotropic with $\vp^2=0$ for $r<a$. This assumption allows for an analytical treatment and is justified when looking at an isolated defect over large distances compared to the characteristic length $L$: since the total active force diverges with system size, the active force that would result from the weak nematic order inside the finite core region is negligible compared to the active force from the bulk region in a large system. Indeed, recent work show that, for isolated defects, the self-advection velocity given by eq.~\eqref{eq_vdefect} is valid even when considering the effect of the weak nematic order inside the core~\cite{Ronning2022}.

Let us call $\mathbf{f} = -f \ve_x$ the external force applied to the core region necessary to stall the defect. We hypothesize further that, in the core region $r<a$, the external force-density has the same mathematical form as the active force in eq.~\eqref{eq_FB_vector} to allow for a similar treatment. Force balance then reads:
\begin{align}
\eta \Delta \mathbf{v} - \mathbf{\nabla} P - \xi \mathbf{v} - \frac{\zeta \Delta \mu}{2r}\ve_x &= 0 \qquad r>a\label{eq_FB_outcore} \\
\eta \Delta \mathbf{v} - \mathbf{\nabla} P - \xi \mathbf{v} - \frac{f}{2\pi a r} \ve_x &= 0 \qquad r<a\ .\label{eq_FB_incore}
\end{align}
At the boundary $r=a$ between these two regions, the global flow and stress fields are continuous. This boundary condition leads to an analytical treatment under the hypothesis that the radius $a$ of the core region is much smaller than the hydrodynamic screening length $L=\sqrt{\eta/\xi}$, as detailed in \ref{app_stalling_wo}. To leading order in $a/L$, we obtain
\begin{equation}\label{eq_pin_force}
f = \sqrt{\frac{\eta}{\xi}}  \frac{\pi^2 \zeta \Delta \mu}{\log\left(a \sqrt{\xi/\eta}\right)}\ .
\end{equation}

A simple way to understand the result of eq.~\eqref{eq_pin_force} is to consider the defect velocity in an active nematic submitted to a force localized at the core of the defect. As the hydrodynamic equations are linear, the velocity is the sum of two contributions: an active contribution $\vv_0$ proportional to the active stress $\zeta \Delta \mu$ and a passive contribution proportional to the applied force $\mathbf{f}$:
\begin{equation}
\vv_{\rm defect}=\vv_0 + \chi \mathbf{f}\ ,
\end{equation}
where $\chi$ is the two-dimensional mobility of the defect. To linear order, this mobility is that of a defect in a passive nematic. A naive guess is given by the Saffman-Delbrück result for the mobility of a disk of radius $a$ in a two-dimensional fluid of shear viscosity $\eta$: $1/\chi \sim \eta/\log(L/a)$, where the large- and short-scale cutoffs are the screening length $L$ and the defect core size $a$, respectively.
Using the active self-advection velocity given by eq.~\eqref{eq_vdefect} leads to eq.~\eqref{eq_pin_force}, up to a numerical prefactor.

\subsection{Effect of cell division/extrusion on the self-advection of a $+1/2$ defect}\label{sec_division}

We now look at the role of the non-conservation of the cell number on the active flow created by a topological defect. This non-conservation stems from cell-division and cell-death processes, as well as from cellular extrusion from the monolayer. We introduce an effective cell-proliferation rate $k$,  which accounts for these three processes. With incompressible cells, the continuity equation then reads:
\begin{equation}\label{eq_continuity}
\nabla \cdot \vv = k\ .
\end{equation}
There is evidence that the effective cell-proliferation rate is influenced by the cellular mechanical environment, and particularly by tissue pressure~\cite{basan2009,montel2011,delarue2013,delarue2014}. At a specific pressure called the homeostatic pressure $P_{\rm h}$, cell divisions and cell deaths and extrusions balance on average, such that $k(P_{\rm h})=0$. Far from the defect, the tissue is in  its homeostatic state $P(r=\infty)=P_{\rm h}$. In the vicinity of the homeostatic state, the pressure-dependent division rate reads to linear order:
\begin{equation}\label{eq_division_rate_pressure}
k(P) = -\frac{1}{\kappa}(P-P_{\rm h})\ ,
\end{equation}
where $\kappa$ is a phenomenological coefficient. It is positive to ensure the monolayer stability and can be interpreted as an effective, long-term, bulk viscosity. 

\begin{figure*}
    \centering
    \includegraphics[width=.9\textwidth]{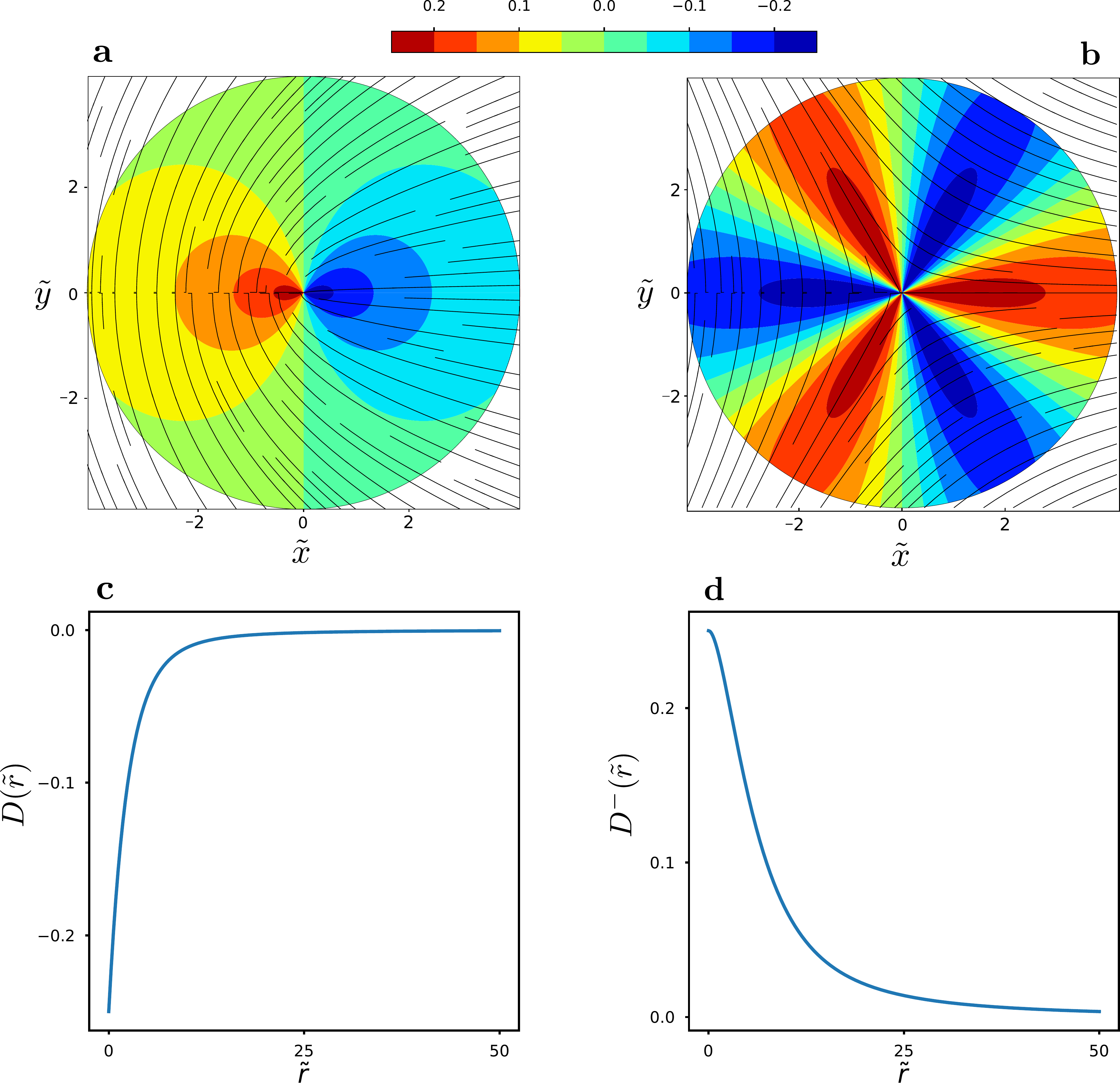}
    \caption{\textbf{a,b.} Colormap of the dimensionless divergence of the velocity---or equivalently of the net division rate $k$---around a $+1/2$ (\textbf{a}) and a $-1/2$ (\textbf{b}) defect. The director-field orientation is represented by solid, thin lines. In the contractile case represented here ($\zeta<0$), there is an asymmetric divergence profile for the $+1/2$ defect, with a positive divergence (orange) at the head and a negative one (blue) at the tail. For the $-1/2$ defect, there is a three-fold symmetry alternating between a positive (orange) and a negative (blue) divergence around the defect.
    \textbf{c,d.} Radial dependence of the dimensionless divergence for a $+1/2$ (\textbf{c}, $\nabla \cdot \tilde{\vv} = D(\tilde r) \cos \theta$) and a $-1/2$ (\textbf{d}, $\nabla \cdot \tilde{\vv}^- = D^-(\tilde r) \cos 3\theta$) defect. The amplitude of this quantity is maximum at the core of the defect and decays with the distance to the core.
}
\label{fig_div_rate}
\end{figure*}

To solve the force-balance equation~\eqref{eq_FB_vector} in this context, we write the velocity field as a sum of a curl-free part and a divergence-free part, also known as the Helmholtz decomposition: $\mathbf{v} = \mathbf{\nabla} \times (\psi\ \ve_z)+\mathbf{\nabla} \phi$. Taking the curl and the divergence of eq.~\eqref{eq_FB_vector} leads to:
\begin{align}
\Delta \left[\Delta \tilde{\psi}  - \tilde{\psi}\right] &= -s\, \frac{\sin \theta}{\tilde{r}^2}\label{eq_FB_psi} \\
\Delta \left[\Delta \tilde{\phi} - \frac{\eta}{\eta+\kappa}\tilde{\phi} \right] &= - s\, \frac{\eta}{\eta+\kappa} \frac{\cos{\theta}}{ \tilde{r}^2}\label{eq_FB_phi}\ ,
\end{align}
using the same adimensional variables and functions as in Section~\ref{sec_self_advection_simple}. These equations are solved in~\ref{app_division} in a similar way as in sec.~\ref{sec_self_advection_simple} to obtain the velocity field. The result for the self-advection velocity of the defect reads:
\begin{equation}\label{eq_vdefect_division}
\mathbf{v}_0 = - \frac{\pi}{4}  \left( 1 + \sqrt{\frac{\eta}{\eta + \kappa}} \right)\times\frac{\zeta \Delta \mu}{2\sqrt{\xi\eta}}\ \ve_x\ .
\end{equation}
The pressure-dependent division rate promotes motion of the defect in the same direction as what is dictated by the active force. 
We recover the incompressible limit of eq.~\eqref{eq_vdefect} in the case of an infinite $\kappa$, that is when the cell number is conserved. In the other limit $\kappa=0$, corresponding to an infinite response of the effective proliferation rate to pressure variations, the velocity of the defect is multiplied by two. In a contractile nematic, active forces associated with a $+1/2$ defect create a low-pressure environment at the head of the defect, which promotes cell division, and the opposite situation at the tail, as illustrated in fig.~\ref{fig_div_rate}a.
At large distances, the divergence of the velocity field vanishes, as illustrated in fig.~\ref{fig_div_rate}c. This pattern of division and death/extrusion further induces motion form head to tail, similarly to the active force. For extensile systems, the situation is reversed. For a $-1/2$ defect, the pattern of effective cell divisions follows a three-fold symmetry. Corresponding expressions are given in~\ref{app_m12_division} and illustrated in fig.~\ref{fig_div_rate}b,d.
Interestingly, such a divergence profile has been observed in collective self-organizations of bacteria~\cite{copenhagen2021}.

\section{Coupling between the velocity field and orientation gradients of the nematic director}\label{sec_rotational_dyn}

All calculations in section~\ref{sec_flow_simple} are performed in the limit of vanishing rotational viscosity $\gamma$. In this limit, the orientation of the director is obtained by minimizing the free energy eq.~\eqref{eq_Frank_energy}. Consequently, the velocity field was computed with the fixed director orientation $\varphi = \theta/2$, corresponding to the orientation field of a passive defect at equilibrium. We now relax this hypothesis to look at the coupling between the flow field and the orientation field of the director. We therefore expect that activity influences the orientation of the director, further modifying the flow field with respect to the results of section~\ref{sec_flow_simple}. Since no global analytical solution can be found to this new system of equations, we compute the velocity field in perturbation in activity. We limit our investigation to first-order terms in activity. Since the flow field is already of order one in activity with the passive defect orientation $\varphi = \theta/2$, the first-order modifications of the orientation contribute only to a second-order term in this velocity field. The first-order contribution of activity to the velocity field can therefore be computed using the passive defect orientation $\varphi = \theta/2$.

As a first approximation, we consider the case where the flow-alignment parameter $\nu$ vanishes, and we further neglect the Ericksen stress tensor $\bm{\Gs}^{\rm E}$. The passive contribution to the stress tensor then reads
\begin{equation} \label{eq_stress_nu0}
\bm{\Gs}^{\rm p} = 2\eta\,\tilde\vu + \frac{1}{2}\left(\vh\otimes\vp - \vp\otimes\vh\right) - P\,\mathbf{1}\ ,
\end{equation}
and the molecular field $\vh$, given by eq.~\eqref{eq_dyn_p}, reads at steady state:
\begin{equation}\label{eq_hmolecular}
\vh = \gamma\left[ (\vv\cdot\nabla)\vp + \bm{\omega}\cdot\vp\right]\, .
\end{equation}

\subsection{Flow field and self-advection velocity}\label{sec_self_advection_gamma}

The active and passive contributions to the stress tensor as given by eqs.~\eqref{eq_stess_active} and \eqref{eq_stress_nu0} lead to the force-balance condition:
\begin{equation} \label{eq_FB_gamma} 
\eta \Delta \mathbf{v} +  \frac{1}{2} \mathbf{\nabla} \times (h_\perp \ve_z) - \mathbf{\nabla} P - \xi (\mathbf{v}+\mathbf{v_0}) - \frac{\zeta \Delta \mu}{2r} \ve_x = 0\ .
\end{equation}
The perpendicular component of the molecular field $h_\perp$ reads, using the equilibrium orientational field $\varphi = \theta/2$:
\begin{equation}\label{eq_hperp}
h_\perp = \frac{\gamma}{2} \left(\frac{v_\theta}{r} - \omega_{\rm s} \right)\ ,
\end{equation}
where $\omega_{\rm s}=(\nabla\times\vv)\cdot\ve_z$ is the vorticity. With this expression, the curl of the force balance equation~\eqref{eq_FB_gamma} leads to:
\begin{equation}\label{eq_psi_4th_order_gamma}
\Delta \left[\Delta \bar{\psi}  - \frac{\lambda}{\bar{r}}\partial_{\bar{r}}\bar{\psi} - \bar{\psi}\right] = -s\frac{\sin \theta}{\bar{r}^2}\, ,
\end{equation}
where we have adimensionalized the equation using the same procedure as in sec.~\ref{sec_self_advection_simple}, except with an effective viscosity $\bar{\eta} = \eta + \gamma/4$, and where $\lambda = \gamma/(4 \bar\eta)$. More precisely, we have defined a new lengthscale $\bar{L}=\sqrt{\bar{\eta}/\xi}$ and a new timescale $\bar{\tau}=2\bar{\eta}/(|\zeta|\Delta\mu)$, defining the dimensionless variable $\bar{r}=r/\bar{L}$ and the dimensionless field $\bar{\psi}=(\bar\tau/\bar L^2)\psi$.
The term inside the brackets of the Laplace operator in eq.~\eqref{eq_psi_4th_order_gamma} is a transformed version of the modified Bessel equation~\cite{bowman1958}.
Similarly to what was done in sec.~\ref{sec_self_advection_simple} and as detailed in~\ref{app_velocity_with}, we determine the velocity field. The resulting self-advection velocity reads:
\begin{equation}\label{eq_velocity_defect_gamma}
\mathbf{v_0} = - c(\lambda) \times\frac{\zeta \Delta \mu}{2\sqrt{\xi \eta}}\ve_x\ ,
\end{equation}
where
\begin{equation}\label{eq_velocity_constant}
c(\lambda) = \frac{\Gamma\left[
        1-\beta/2 \right]\ \Gamma\left[ 1+1/(2\beta)\right]}{2\sqrt{\lambda}\ \Gamma\left[ 3/2-\beta/2 \right]\ \Gamma\left[ 3/2+1/(2\beta)\right]}
\end{equation}
with  $\Gamma$ the Euler gamma function~\cite{abramowitz1964} and $\beta = \sqrt{1+\lambda^2/4} + \lambda/2$.

\begin{figure}[htb]
    \centering
    \includegraphics[width=0.4\textwidth]{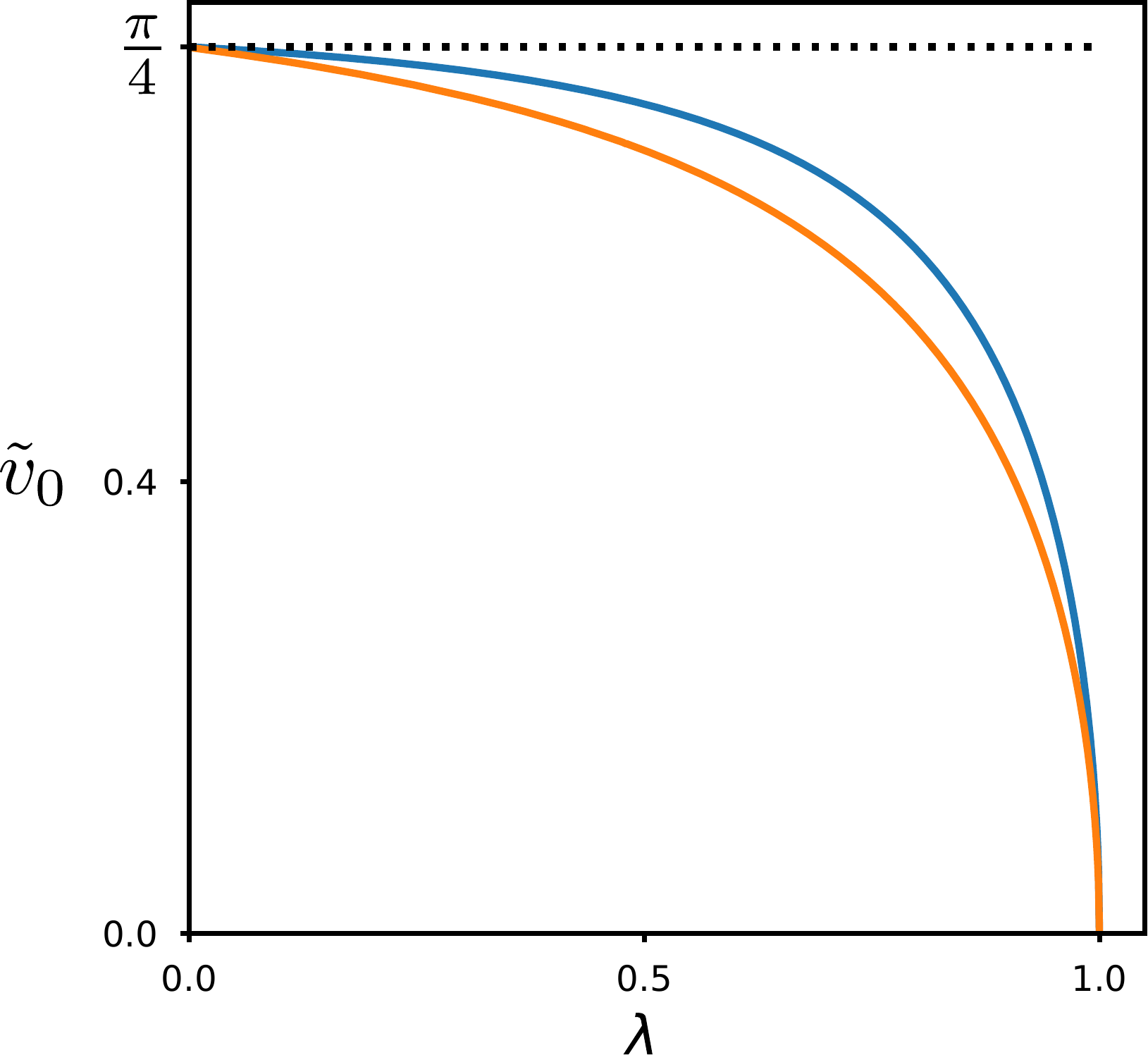}
    \caption{Self-advection velocity of a $+1/2$ defect in units of $\zeta \Delta \mu/(2\sqrt{\xi\eta})$ and as a function of $\lambda$. The dotted horizontal line shows the value $\pi/4$, corresponding to eq.~\eqref{eq_vdefect} when $\lambda=0$. Two cases are presented: without the Ericksen contribution to the stress (blue curve), as determined by eq.~\eqref{eq_velocity_defect_gamma}, and with the Ericksen contribution (orange curve), determined numerically (see section~\ref{sec_ericksen}).}
    \label{fig_velocity_ericksen}
\end{figure}

We plot in fig.~\ref{fig_velocity_ericksen} the dimensionless amplitude $\tilde{v}_0$ of the self-advection velocity as a function of $\lambda$, both not considering and considering the Ericksen stress (see section~\ref{sec_ericksen}).
The rotational viscosity overall decreases this amplitude, which is expected since it corresponds to an additive source of dissipation.

\subsection{Stall force}\label{sec_pinning_gamma}

Proceeding similarly as in section~\ref{sec_pinning_simple}, the in-core force balance is unchanged as compared with eq.~\eqref{eq_FB_incore}, since there is no nematic order within the core region, and the out-core force balance condition is given by eq.~\eqref{eq_FB_gamma}. Imposing continuity of the flow and stress fields at the boundary $r = a$ between the in-core and out-core regions as detailed in \ref{app_stalling_with} gives, to leading order in $a/\bar{L}$:
\begin{equation}\label{eq_pin_force_gamma}
f = - \sqrt{\frac{\bar{\eta}}{\xi}}\ \zeta\Delta\mu\ g(\lambda) \left(a \sqrt{\frac{\xi}{\bar{\eta}}} \right)^{\beta -1}\ ,
\end{equation}
where
\begin{align}\label{eq_glambda}
g(\lambda) =& 2^{\beta-\lambda}\ \Gamma\left[1-\beta/2\right]\ \Gamma\left[ 1+1/(2\beta)\right]\nonumber\\
&\times 6\pi(2-\lambda)\frac{\left[ \lambda(4-\lambda(1+\beta))+24(\beta-\lambda/2-1)\right]}{(48-20\lambda^2+\lambda^4)\Gamma(1+\beta-\lambda/2)}\,.
\end{align}
The force vanishes as a power law of the core size $a$ with an exponent $\beta-1$, which varies between 0 and $(\sqrt{5}-1)/2$ as $\lambda$ varies between 0 and 1.
Compared to the previous case $\gamma=0$, the stall force here is comparatively smaller in the limit of a small core size compared to the hydrodynamic length $\bar{L}$, with a power-law dependence on the core size $a$ rather than a logarithmic one.

\subsection{First-order active correction to the director orientation}\label{sec_orientation}

We can now investigate the effect of the flow on the orientation of the director, to first order in activity. With a non-vanishing rotational viscosity $\gamma$, the perpendicular component of the molecular field is given by 
eq.~\eqref{eq_hperp}. Given the definition of the molecular field and the expression of the free energy eq.~\eqref{eq_Frank_energy}, we obtain, in terms of the stream function $\psi$,
\begin{equation}\label{eq_DeltaPhi}
K \Delta\varphi = \frac{\gamma}{2}\left[\Delta \psi - \frac{\partial_r \psi}{r} \right]\ .
\end{equation}
For reasons similar to those presented in sec.~\ref{sec_self_advection_simple}, the stream function remains of the form $\psi(r,\theta) = \psi(r) \sin \theta$, and we obtain a similar form for the term proportional to the rotational viscosity $\gamma$ in the expression of $\varphi(r,\theta)$:
\begin{align}\label{eq_phi_sol}
& \varphi(r,\theta) = \frac{\theta}{2} + \frac{\gamma}{2K}  \left\{r v_r(r) \right.\nonumber \\
& \quad + \left. \frac{1}{2} \left( r \int_0^r{\frac{v_\theta(u)}{u}\: \diff u} - \frac{1}{r} \int_0^r{u v_\theta(u) \: \diff u}\right) \right\}\sin \theta\ .
\end{align}

Let us now investigate the behavior of the correction $\delta\varphi(r,\theta)$ to the passive orientation $\varphi^0(r,\theta) = \theta/2$ at large distances $r \gg \bar L$. Asymptotically, we get:
\begin{equation}\label{eq_phi_far}
\delta\varphi(r,\theta) \underset{r \gg \bar L}{\approx} \frac{\gamma}{4K} r \ln{\left(\frac{r}{r_0}\right)}v_0 \, ,
\end{equation}
where $r_0$ is an arbitrary length that has been introduced based on dimensional analysis. Since this expression diverges with $r$, the solution given by eq.~\eqref{eq_phi_sol} is only valid up to a characteristic lengthscale $L_{\text{defect}}\propto K/(\gamma v_0)$. Using the scaling of eq.~\eqref{eq_velocity_defect_gamma}, we define
\begin{equation}\label{eq_L_defect}
L_{\text{defect}} = \frac{\sqrt{\xi(\eta+\gamma/4)}}{\gamma/4}\ \frac{K}{|\zeta \Delta \mu|}=\frac{L_a^2}{\lambda \bar{L}}\, ,
\end{equation}
where  $L_{\rm a} = \sqrt{K/|\zeta \Delta \mu|}$ is the active length, corresponding to the scale above which uniform bands of active nematics become unstable~\cite{voituriez2005,edwards2009} and active turbulence is expected~\cite{giomi2015,doostmohammadi2018}. This result gives an insight into the interplay between active and hydrodynamic stresses. Note that our description is only valid if $L_{\rm a} \gg \bar{L}$. In this limit, eq.~\eqref{eq_phi_far} is valid at distances $r$ from the core region such that $\bar L\ll r\ll L_{\rm a}$. If this is not the case, the homogeneous nematic system does not have a stable steady state and the passive orientation $\varphi^0(r,\theta) = \theta/2$ is spontaneously destroyed over length scales of order $\bar{L}$.

\subsection{Contribution of the Ericksen stress tensor to the self-advection velocity}\label{sec_ericksen}

Until now, we have only considered the coupling between the velocity field and the nematic order through the torques associated with the antisymmetric component of the stress tensor. In the expression of the stress tensor, we have however neglected the Ericksen stress tensor $\bm{\Gs}^{\rm E}$ as well as the flow-alignment coupling term proportional to the parameter $\nu$. In perturbation theory around a homogeneous steady state, neglecting the Ericksen stress is in general justified by the fact that it is second order in the gradients of the director orientation, and therefore nonlinear in activity~\cite{degennes-prost}. Here, however, the Ericksen stress is relevant because the ground state upon which activity sets in has a non-homogeneous director orientation $\varphi=\theta/2$. The force density resulting from the Ericksen stress is given by the Gibbs-Duhem relation~\eqref{eq_Gibbs_Duhem}, which here reads
\begin{equation}\label{eq_Ericksen_defect}
\nabla \cdot \bm{\Gs}^{\rm E} =-\frac{h_\perp}{2r}\ \ve_\theta\, .
\end{equation}
Accounting for the Ericksen stress while keeping $\nu=0$, the curl of the force-balance condition leads to
\begin{equation}\label{eq_app_psi_ericksen}
\Delta \left[\Delta \bar{\psi}  - \frac{\lambda}{\bar{r}}\partial_{\bar{r}}\bar{\psi} - \bar{\psi}\right] + \frac{\lambda}{\bar{r}} \partial_{\bar{r}}\left[ \Delta\bar{\psi} - \frac{\partial_{\bar{r}} \bar{\psi} }{r} \right]= -s\frac{\sin \theta}{\bar{r}^2}\, ,
\end{equation}
with the same dimensionless units as in eq.~\eqref{eq_psi_4th_order_gamma}. Contrary to the previous equations for the stream function in this paper, this equation has no direct analytical solution. Using asymptotic expansions close to the core and at large distances, we can however determine the velocity of the defect using a shooting method. We plot in fig.~\ref{fig_velocity_ericksen} the resulting self-advection velocity of the defect as a function of $\lambda$ (orange curve), in comparison to the analytical solution obtained in the absence of Ericksen stress (blue curve): there is a decrease in amplitude due to the Ericksen stress, which is expected given that it represents an additional source of rigidity.

\section{Discussion}\label{sec_discussion}

In this manuscript, we have studied the flows generated by $\pm1/2$ topological defects of an active nematic system in two dimensions, which interacts with its underlying substrate via viscous drag. The active nematic system lies deep in the nematic phase, with an order parameter of constant, maximal amplitude. Our study is inspired by monolayers of elongated cells at confluence. In the case of a vanishing rotational viscosity, we derived analytical expressions for the flow field. In the case of a $+1/2$ topological defect, we determined its self-advection velocity and stall force. We showed that the non-conservation of cell number leads to an increase in the self-advection velocity. We then investigated the effect of the rotational viscosity, which couples the velocity field to the nematic order. Taking the flow-alignment parameter to zero and in the absence of the Ericksen contribution to the stress, we derived analytical expressions of the self-advection velocity and stall force, making use of the passive, equilibrium configuration of the director around the defect. We finally studied the validity domain of the small departure from the equilibrium nematic-order configuration, as well as the influence of the Ericksen stress.

Our results show that the self-advection velocity is linear in activity for a $+1/2$ defect, as it should, as well as its corresponding stall force. The self-advection velocity of a $+1/2$ defect decreases in the presence of a non-vanishing rotational viscosity, which adds an extra source of dissipation. Accordingly, the stall force is also smaller, with a qualitatively different dependence on the size of the core region of the defect, where the nematic order vanishes. On the contrary, the non-conservation of the cell number increases the self-advection velocity, when coupled linearly to the departure from a homeostatic isotropic stress. Finally, accounting for the Ericksen stress lowers even further that velocity, with a similar qualitative interpretation.

Several assumptions used in our study can be questioned. A first restriction consisted in studying
an isolated, topological defect in an infinite domain, as defects in an actual, free
monolayers come into pairs to ensure a vanishing total topological charge. This
assumption allowed for an analytical treatment that does not depend on finite-domain boundary
conditions. Importantly, section~\ref{sec_orientation} provided a characteristic system size, within which our calculations are valid, and beyond which active effects drive the nematic far from its passive, equilibrium configuration. At a qualitative level, the presence of other defects introduces another length scale in the problem, namely the characteristic distance $d$ between defects. If this distance is such that $d\gg \bar L$, our results remain valid as the defects interact only very weakly. The interactions between defects becomes relevant in the limit where $d\lesssim \bar L$. In this case, one should consider the nematic layer as a gas of interacting defects~\cite{shankar2018}. 
Within our approach, computing the flow created by a pair of $+1/2$ and $-1/2$ defects is the next logical step.
A second restriction was to consider only partially the coupling between flow and orientation. In particular, for a non-vanishing rotational viscosity, we have limited our study to the case of a vanishing flow-alignment parameter. Beyond this approximation, no analytical solutions to the flow equations could be found.
Finally, considering proliferation and extrusion is of prominent relevance for living systems, and we provided a first computation
of this effect on the self-advection velocity of $+1/2$ defects.

The initial goal of this work was to explain the observation of Sarkar
\emph{et al.} in ref \cite{sarkar2021} of motionless defects that are
preferential sites for multilayer formation. We computed
the self-advection velocity of motile defects and the force necessary to stall them. However, the origin of the stalling and the mechanism by which another layer of cells forms at the defect is still elusive. Accounting for pressure-dependent proliferation shows that, on average, more extrusion is to be
expected compared to the rest of the monolayer at the tail of $+1/2$, contractile
defects.
Recent studies by Vafa \& Mahadevan~\cite{vafa2021} and by
Hoffman \emph{et al.}~\cite{hoffmann2021} show that, when considering a
deformable surface, there is an out-of-plane force at topological defects
that could be responsible for extrusion and multilayering.
\begin{acknowledgements}
We thank P. Silberzan and T. Sarkar from Institut Curie for sharing their experimental results with us and useful discussions. LB received a PhD fellowship from the doctoral school Physique en Ile-de-France (EDPIF) and support by the Coll\`ege de France foundation and Institut Curie. This work received support from the grants ANR-11-LABX-0038, ANR-10-IDEX-0001-02.
\end{acknowledgements}


\bibliographystyle{ieeetr}

\appendix
\section{Modified Bessel functions}\label{app_bessel}

\subsection{Definition}

The modified Bessel functions $I_\alpha,K_\alpha$ are the two general solutions of the following equation:
\begin{equation}\label{eq_app_bessel}
r^2 f''(r) + r f'(r) - (\alpha^2+r^2) f(r) = 0\ .
\end{equation}
A related equation to the modified Bessel equation~\eqref{eq_app_bessel} is given by Bowman~\cite{bowman1958}:
\begin{equation}\label{eq_app_bessel_bowman}
r^2 f''(r) + (1-2a) r f'(r) - (b^2 c^2 r^{2c} - a^2 + d^2 c^2) f(r) = 0\ .
\end{equation}
Two independent solutions of eq.~\eqref{eq_app_bessel_bowman} are $r^a I_d(b r^c)$ and $r^a K_d(b r^c)$.

\subsection{Derivatives}\label{app_bessel_derivatives}

The first-order derivatives of the modified Bessel functions are given by:
\begin{align}
    I_\alpha'(r) &= I_{\alpha-1}(r)-\frac{\alpha}{r}I_\alpha
    \label{eq:app_Ider} \\
    K_\alpha'(r) &= -\left(K_{\alpha-1}(r)+\frac{\alpha}{r}K_\alpha\right)
    \label{eq:app_Kder}\\
    I_0'(r) &= I_1(r) \label{eq:app_I0diff}\\
    K_0'(r) &= -K_1(r) \label{eq:app_K0diff}
\end{align}

\subsection{Asymptotic expansions}\label{app:bessel_asymptotics}

We also use the following asymptotic expansions:
\begin{align}
    I_\alpha(r) &\underset{r \rightarrow
    0}{\simeq}\frac{1}{\Gamma(\alpha+1)}\left( \frac{r}{2} \right)^\alpha
    \label{app_asympt_bessel-I0}\\
    K_\alpha(r) &\underset{r \rightarrow
    0}{\simeq}
    \begin{cases}
        \frac{\Gamma(\alpha)}{2} \left( \frac{2}{r} \right)^\alpha \quad
        \text{if } \alpha>0 \\
        -\ln (\frac{r}{2}) - \gamma \quad
        \text{if } \alpha=0
    \end{cases}
     \label{app_asympt_bessel-K0}
\end{align}
where $\gamma$ denotes here the Euler's constant. At infinity, we use
\begin{align}
    I_\alpha(r) &\underset{r \rightarrow
    +\infty}{\simeq} \frac{e^r}{\sqrt{2\pi r}} \label{eq:asympt_bessel-Iinf}\\
    K_\alpha(r) &\underset{r \rightarrow
    +\infty}{\simeq} \sqrt{\frac{\pi}{2r}} e^{-r}
    \label{eq:asympt_bessel-Kinf}
\end{align}

\section{Velocity field}\label{app_velocity_comp}

\subsection{Limit of a vanishing rotational viscosity}\label{app_velocity_wo}

This appendix is dedicated to the full computation of the velocity
starting from eq.~\eqref{eq_psi_4th_order}.
As justified in the main text, the stream function reads $\tilde{\psi}(\tilde r,\theta) = \tilde{\psi}(\tilde r) \sin \theta$.
Integrating one Laplace operator gives:
\begin{equation}
\Delta \tilde{\psi}(\tilde r,\theta) - \tilde{\psi}(\tilde r,\theta) = \left[ A \tilde r + \frac{B}{\tilde r} + s \right] \sin \theta\ ,
\end{equation}
where $A$ and $B$ are integration constants. The radial dependence of the stream function then satisfies:
\begin{equation}\label{eq_psi_dev}
\frac{\diff^2 \tilde{\psi}(r)}{\diff \tilde r^2} + \frac{1}{\tilde r}\frac{\diff \tilde{\psi}(\tilde r)}{\diff \tilde r} - \left(1+\frac{1}{\tilde r^2}\right)\tilde{\psi}(r) = A \tilde r + \frac{B}{\tilde r} + s\ .
\end{equation}
The homogeneous solution $\tilde{\psi}^0$ to this equation reads:
\begin{equation}\label{eq_solhomo_psi}
    \tilde{\psi}^0(\tilde r) = A_0 I_1(\tilde r) + B_0 K_1(\tilde r)\ ,
\end{equation}
where $A_0$ and $B_0$ are integration constants. The Wronskian associated to eq.~\eqref{eq_psi_dev} reads
\begin{equation}\label{eq_app_Wronskian}
I_1(\tilde r) K'_1(\tilde r) - I'_1(\tilde r) K_1(\tilde r) = -\frac{1}{\tilde r}\ ,
\end{equation}
which leads to:
\begin{align}\label{eq_sol_psi_full}
    \tilde\psi(\tilde r) &=\left\{ I_1(\tilde r)\left( A_0 - \int_{\tilde r}^{+\infty}{K_1(u)(Au^2 + su + B)\:
        \diff u} \right) \right. \nonumber\\
        &+\left. K_1(\tilde r) \left( B_0 
        - \int_0^{\tilde r}{I_1(u)(Au^2 + su + B)\: \diff u} \right) \right\} \ .
\end{align}

The velocity field $\mathbf{\tilde v} = \tilde v_r(\tilde r) \cos \theta \: \mathbf{e_r} + \tilde v_\theta(\tilde r) \sin \theta
\: \mathbf{e_\theta}$ is obtained from the derivatives of the stream function: $\tilde v_r(\tilde r) = \tilde \psi(\tilde r)/ \tilde r$ and $\tilde v_\theta(\tilde r) = -\diff \tilde \psi(\tilde r)/\diff \tilde r$. We obtain
\begin{align}
    \tilde{v}_r(\tilde r) &= \tilde r^{-1} \left[I_1(\tilde r)
    \left( A_0 - \int_{\tilde r}^{+\infty}{K_1(u)(Au^2 +  su + B)\:
        \diff u} \right)  \right. \nonumber\\
         &\qquad \qquad \quad+\left. K_1(\tilde r) \left( B_0 
        - \int_0^{\tilde r}{I_1(u)(Au^2 + su + B)\: \diff u} \right) \right]
        \label{eq:app_vr_full}\\
    \tilde{v}_\theta(\tilde r) &= -\left[\left( 
        I_0(\tilde r)-\frac{I_1(\tilde r)}{\tilde r} \right)
    \left( A_0 - \int_{\tilde r}^{+\infty}{K_1(u)(Au^2 +  su + B)\:
        \diff u} \right)  \right. \nonumber\\
         &\qquad \quad-\left. \left(K_0(\tilde r) + \frac{K_1(\tilde r)}{\tilde r} \right) \left( B_0 
        - \int_0^{\tilde r}{I_1(u)(Au^2 + su + B)\: \diff u} \right) \right]\ .
    \label{eq:app_vt_full}
\end{align}
Imposing a finite velocity at the origin leads to $B_0 = 0$ and $B=0$, as a finite velocity at infinity leads to $A_0=0$. We then obtain $\tilde{v}_r(\tilde r=0)=-(A + s\pi/4)$ and $\tilde{v}_\theta(\tilde r=0)= A + s\pi/4$ at the origin, and $\tilde{v}_r(\tilde r=\infty)= -A$ and $\tilde{v}_\theta (\tilde r=\infty)= A$ at infinity. Imposing $\tilde v_r(0) = \tilde v_\theta(0) = 0$ yields $A=-s\pi/4$, and imposing $\tilde v_r(\tilde r=\infty) = -\tilde v_0$ and $\tilde v_\theta(\tilde r=\infty) = \tilde v_0$ yields $A = \tilde{v}_0$. Coming back to the physical units, these lead to the self-advection velocity given by eq.~\eqref{eq_vdefect}.

\subsection{Velocity field with a finite rotational viscosity}\label{app_velocity_with}

We start the computation from eq.~\eqref{eq_psi_4th_order_gamma} with the dependence $\tilde{\psi}(\tilde r,\theta) =
\tilde{\psi}(\tilde r) \sin \theta$. Integrating one
Laplace operator gives:
\begin{equation}\label{eq_app_psi_gamma}
    \frac{\diff^2 \tilde{\psi}(r)}{\diff \tilde r^2}  + \frac{(1-\lambda)}{\tilde r}\frac{\diff \tilde{\psi}(r)}{\diff \tilde r} 
    - (1+\frac{1}{\tilde r^2}) \tilde{\psi} =A\:\tilde r +
    \frac{B}{\tilde r} + s \ .
\end{equation}
The left-hand side of this equation is of the form~\eqref{eq_app_bessel_bowman} with $a = \lambda/2$, $b=1$, $c=1$, and
$d=\sqrt{1+\lambda^2/4}$. Following~\ref{app_bessel}, two independent homogeneous solutions to eq.~\eqref{eq_app_psi_gamma} are
$\tilde r^{\lambda/2}I_{\alpha}(\tilde r),\tilde r^{\lambda/2}K_{\alpha}(\tilde r)$, with $\alpha = \sqrt{1+\lambda^2/4}$.
Following a similar procedure as in~\ref{app_velocity_wo}, the dimensionless velocity field has the form:
\begin{align}
    \tilde{v}_r =  \tilde r^{\lambda/2-1} &\left\{ I_\alpha(\tilde r) \left( A_0 -
        \int_{\tilde r}^{+\infty}{u^{-\lambda/2}K_\alpha(u)(Au^2 +  su + B)\:
    \diff u}\right)\right. \nonumber \\
    &+\left. K_\alpha(\tilde r) \left( B_0 -
    \int_0^{\tilde r}{u^{-\lambda/2}I_\alpha(u)(Au^2 +  su + B)\: \diff u}\right)
\right\} \nonumber\\
    &\times \cos \theta \label{eq:ch3_speed_full_gamma-r} \\
    \tilde{v}_\theta =  \tilde r^{\lambda/2}
    &\left\{ \left( \frac{\alpha - \lambda/2}{\tilde r} I_\alpha(\tilde r) - I_{\alpha-1}(\tilde r) \right) \times \right. \nonumber \\
    & \left. \quad \left( A_0 -
        \int_{\tilde r}^{+\infty}{u^{-\lambda/2}K_\alpha(u)(Au^2 +  su + B)\:
    \diff u}\right) \right. \nonumber \\
    & + \left.\left( \frac{\alpha - \lambda/2}{\tilde r} K_\alpha(\tilde r) + K_{\alpha-1}(\tilde r)\right)\right.\times \nonumber \\
    & \left. \quad \left( B_0 -
    \int_0^{\tilde r}{u^{-\lambda/2}I_\alpha(u)(Au^2 + s u + B)\: \diff u}\right)
    \right\} \sin \theta \label{eq:ch3_speed_full_gamma-t} \ .
\end{align}
The boundary conditions still set $A_0=0$, $B_0=0$, $B=0$, and $A=\tilde{v}_0$,  similarly as in~\ref{app_velocity_wo}. The asymptotic expansion of the velocity field close to the core reads:
\begin{align}\label{eq_app_velocityr_zero_gamma}
    \tilde{v}_r(\tilde r) &\underset{\tilde r\ll 1}{\simeq} 
    -\frac{2^{-\alpha}}{\Gamma(\alpha+1)}(sC_1^\lambda  + \tilde{v}_0 C_2^\lambda )
    \tilde r^{\lambda/2+\alpha-1} \\
    \tilde{v}_\theta(\tilde r) &\underset{\tilde r\ll 1}{\simeq} 
    \frac{2^{-\alpha}(\lambda/2 + \alpha)}{\Gamma(\alpha+1)}
    (sC_1^\lambda  + \tilde{v}_0 C_2^\lambda) \tilde r^{\lambda/2+\alpha-1} \ ,\label{eq:app_velocityt_zero_gamma}
\end{align}
with
\begin{align}
    C_1^\lambda &= \int_0^\infty{\diff u \: u^{1-\lambda/2} K_\alpha(u)}\nonumber\\
   & =2^{- \lambda/2}\Gamma\left[
        1-\frac{1}{2}\left(\alpha+\frac{\lambda}{2}
\right) \right] \: \Gamma\left[ 1+\frac{1}{2}\left(
\alpha - \frac{\lambda}{2}
\right) \right]\label{eq:app_C1}\\
    C_2^\lambda &= \int_0^\infty{\diff u \: u^{2-\lambda/2} K_\alpha(u)}\nonumber\\
    &= 2^{1- \lambda/2}\Gamma\left[ \frac{1}{2}\left(3 - 
\alpha - \frac{\lambda}{2}
\right) \right] \: \Gamma\left[ \frac{1}{2}\left(3 +
\alpha - \frac{\lambda}{2}
\right) \right] \ .
    \label{eq_app_C2}
\end{align}
Contrary to the case of a vanishing rotational viscosity, the velocity field in the defect reference frame vanishes close to the core for any finite value of $\tilde{v}_0$, since the exponent $\lambda/2+\alpha-1$ is positive.
To set the value of $\tilde{v}_0$, we must consider the
tangential stress $\sigma_{r \theta}$. Using the
angular dependence of the stream function and force balance, its dimensionless version reads
\begin{align}\label{eq_app_sigmart_tot_gamma}
    \tilde \sigma_{r \theta}(\tilde r,\theta)=  \left(
    2\frac{\lambda - 1}{\tilde r}(\tilde{v}_\theta(\tilde r)+\tilde{v}_r(\tilde r)) - \tilde r \tilde{v}_r(\tilde r) -
    \tilde{v}_0\:\tilde r \right)\sin\theta\ .
\end{align}
Since $\lambda/2+\alpha-2$ is negative, $\tilde{v}_0$ must equal $-sC_1^\lambda/C_2^\lambda$
for this tangential stress to remain finite at the origin.
This leads to eq.~\eqref{eq_velocity_defect_gamma} in physical units.

\section{Stall force}\label{app_stalling}

\subsection{Limit of vanishing rotational viscosity}
\label{app_stalling_wo}
In the limit of vanishing rotational viscosity, the stall force is determined by the velocity and pressure solutions to eqs.~\eqref{eq_FB_outcore} and \eqref{eq_FB_incore}, together with the incompressibility condition $\nabla \cdot \vv = 0$.
Quantities defined inside the core $r<a$ bear the superscript `c'.
Using the dimensionless units of sec.~\ref{sec_flow_simple}, the stream function satisfies:
\begin{align}
    \Delta \left[\Delta \tilde{\psi}  -
    \tilde{\psi}\right] &= -s\frac{\sin \theta}{\tilde{r}^2} \qquad \tilde r>\tilde a
    \label{app_psi_out_4th}\\ 
    \Delta \left[\Delta \tilde{\psi}^\mathrm{c}  -
    \tilde{\psi}^\mathrm{c}\right] &= -\tilde{f}\frac{\sin \theta}{\tilde{r}^2} \qquad \tilde r<\tilde a\ , 
    \label{app_psi_in_4th}
\end{align}
with $\tilde a=a/L$ and $\tilde{f} = f/(\pi a \zeta \Delta \mu)$ the normalized core radius and overall force
applied to the core.
The solution for the velocity field outside the core is given
by eqs.~\eqref{eq:app_vr_full} and \eqref{eq:app_vt_full} with $A_0=0$ and $A=\tilde{v}_0$ as in~\ref{app_velocity_wo}.
The velocity field inside the core however reads:
\begin{align}
    \tilde{v}_r^\mathrm{c}(\tilde r) &= \tilde r^{-1} \left[I_1(\tilde r)
    \left( A_0^\mathrm{c} + \int_{0}^{\tilde r}{K_1(u)(A^\mathrm{c}u^2 + \tilde{f} u + B^\mathrm{c})\:
        \diff u} \right)  \right. \nonumber\\
         & \qquad+\left. K_1(\tilde r) \left( B^\mathrm{c}_0 
        - \int_0^{\tilde r}{I_1(u)(A^\mathrm{c}u^2 + \tilde{f}u + B^\mathrm{c})\: \diff u} \right) \right]
        \label{eq:app_vr_full_in}\\
    \tilde{v}^\mathrm{c}_\theta(\tilde r) &= -\left[\left( 
        I_0(\tilde r)-\frac{I_1(\tilde r)}{\tilde r} \right) \times \right.\nonumber\\
    & \qquad \quad\left.\left( A^\mathrm{c}_0 + \int_{0}^{\tilde r}{K_1(u)(A^\mathrm{c}u^2 + \tilde{f} u + B^\mathrm{c})\:
        \diff u} \right)  \right. \nonumber\\
         &\quad - \left(K_0(\tilde r) + \frac{K_1(\tilde r)}{\tilde r} \right) \times \nonumber\\
        & \qquad \quad \left.\left( B^\mathrm{c}_0 
        - \int_0^{\tilde r}{I_1(u)(A^\mathrm{c}u^2 + \tilde{f}u + B^\mathrm{c})\: \diff u} \right) \right]\ .
    \label{eq:app_vt_full_in}
\end{align}
The boundary condition $\vv^\mathrm{c}(\tilde r=0)=\mathbf{0}$ imposes that $B^\mathrm{c}$, $B^\mathrm{c}_0$, and $A_0^\mathrm{c}$ vanish.
The other integration
constants are set by imposing the continuity of
the velocity and stress fields at the boundary of the core $r=a$.
Introducing $\tilde{P}$, the pressure normalized by $|\zeta| \Delta \mu/2$ and taking the divergence of the force-balance eqs.~\eqref{eq_FB_outcore} and \eqref{eq_FB_incore}, we obtain:
\begin{align}
    \Delta \tilde{P} &= s\frac{\cos \theta}{\tilde r^2}\qquad \tilde r>\tilde a
    \label{eq:app_pressure_out_diff}\\
    \Delta \tilde{P}^\mathrm{c} &= \tilde{f}\frac{\cos \theta}{\tilde r^2}\qquad \tilde r<\tilde a\ .
    \label{eq:app_pressure_in_diff}
\end{align}
These equations are solved by:
\begin{align}
    \tilde{P} (\tilde r,\theta) &= \left[A_* \tilde r + \frac{B_*}{\tilde r} -
    s \right] \cos \theta \label{eq:app_pressure_in} \\
    \tilde{P}^\mathrm{c} (r,\theta) &= \left[A_*^\mathrm{c} \tilde r + \frac{B_*^\mathrm{c}}{\tilde r} -
         \tilde{f}\right] \cos \theta\ ,
    \label{eq:app_pressure_out}
\end{align}
where $A_*$, $B_*$, $A_*^\mathrm{c}$, and $B_*^\mathrm{c}$ are integration constants. Pressure and velocity are linked by force balance, imposing
\begin{align}
    B_* &= B \label{eq:app_B_FB} \\
    B_*^\mathrm{c} &= B^\mathrm{c} \label{eq:app_Bc_FB} \\
    A_* &= 0
    \label{eq:app_A_FB}\\
    A_*^\mathrm{c} &= A^\mathrm{c} - \tilde{v}_0\ .
    \label{eq:app_Ac_FB}
\end{align}
The normal and tangential components of the stress read $\sigma_{rr}=2\eta \partial_r v_r - P$ and $\sigma_{r\theta}=\eta[(\partial_\theta v_r -v_\theta)/r + \partial_r v_\theta]$, respectively. We impose the continuity of the velocity and stress fields at the boundary of the core in $r=a$ in the limit of a small core size $\tilde a\ll 1$, in which case we can make use of the following asymptotic expressions:
\begin{align}
    \tilde{v}_r^\mathrm{c}(\tilde r) &\underset{\tilde r\ll 1}{\simeq}
    \frac{A^\mathrm{c}}{8} \tilde r^2 + \frac{\tilde f}{3} \tilde r \label{app_vcore_0} \\
    \tilde{v}_r(\tilde r) &\underset{\tilde r\ll 1}{\simeq}
    -s\frac{C^0_1}{2}-\tilde{v}_0\frac{C^0_2}{2}+\frac{B_0}{\tilde r^2}+\frac{B}{2}\log \tilde r \label{app_vr_0}\\
    \tilde{v}_\theta^\mathrm{c}(\tilde r) &\underset{\tilde r\ll 1}{\simeq} -\frac{3A^\mathrm{c}}{8} \tilde r^2 - \frac{2\tilde f}{3} \tilde r \label{app_vtcore_0}\\
    \tilde{v}_\theta(\tilde r) &\underset{\tilde r\ll 1}{\simeq}
    s\frac{C^0_1}{2} + \tilde{v}_0\frac{C^0_2}{2}+\frac{B_0}{\tilde r^2}-\frac{B}{2}\log \tilde r \label{app_vt_0}\ ,
\end{align}
with $C_1^0$ and $C_2^0$ given respectively by eqs.~\eqref{eq:app_C1} and~\eqref{eq_app_C2} with $\lambda=0$.
Finally, we obtain the stall force by imposing
$\tilde{v}_0=0$. The result is given by eq.~\eqref{eq_pin_force}, to leading order in $\tilde{a}=a/L=a\sqrt{\xi/\eta}$.

\subsection{Finite rotational viscosity}\label{app_stalling_with}
The procedure to determine the stall force with a finite rotational viscosity resembles that presented in~\ref{app_stalling_wo}.
Since we assume no nematic order within the core region, the fields inside the core are unchanged as compared with~\ref{app_stalling_wo}. The pressure field outside the core does not depend on the rotational viscosity and is unchanged. The quantity that changes is the velocity field outside the core region, now given by eqs.~\eqref{eq_app_velocityr_zero_gamma}--\eqref{eq_app_C2}.
Close to the core, for $\bar{a} < \bar{r} \ll 1$, the velocity components have the following asymptotic expressions:
\begin{align}
    \bar{v}_r(\bar{r}) \underset{\bar{r}\ll 1}{\simeq}&
    -\frac{2^{-\alpha}}{\Gamma(\alpha+1)}(s\,C^\lambda_1+\tilde{v}_0 \, C^\lambda_2) \bar{r}^{\lambda/2+\alpha-1} \nonumber \\
    & +2^{\alpha-1}B_0 \Gamma(\alpha) \bar{r}^{\lambda/2-\alpha-1} - \frac{B}{\lambda}\label{app_vr_gamma_0}\\
    \bar{v}_\theta(\bar{r}) \underset{\bar{r}\ll 1}{\simeq}&
    \frac{2^{-\alpha}(\lambda/2+\alpha)}{\Gamma(\alpha+1)}(s\,C^\lambda_1+\tilde{v}_0 \, C^\lambda_2) \bar{r}^{\lambda/2+\alpha-1}\nonumber \\
    &+2^{\alpha-1}(\alpha -\lambda/2)B_0 \Gamma(\alpha) \bar{r}^{\lambda/2-\alpha-1} + \frac{B}{\lambda}\ , \label{app_vt_gamma_0}
\end{align}
with the notations of~\ref{app_velocity_with}.
Using the expressions eqs.~\eqref{eq:app_pressure_in},  \eqref{eq:app_pressure_out}, \eqref{app_vcore_0},      \eqref{app_vtcore_0}, \eqref{app_vr_gamma_0}, and \eqref{app_vt_gamma_0}, continuity at the core boundary leads to the stalling force given by eq.~\eqref{eq_pin_force_gamma} for $\tilde{v}_0=0$\footnote{Note that special care must be taken when matching the fields inside and outside the core, because normalizations have been done in these two instances using two different characteristic lengths, $ L = \sqrt{\eta/ \xi}$ and $\bar{L} = \sqrt{(\eta + \gamma/4)/ \xi}$, respectively.}.

\section{Cell division/extrusion}\label{app_division}
In section~\ref{sec_division}, we introduce the Helmholtz decomposition $\mathbf{v} = \mathbf{\nabla} \times (\psi\ \ve_z)+\mathbf{\nabla} \phi$. 
In dimensionless units, $\tilde \phi$ satisfies eq.~\eqref{eq_FB_phi}, which is solved by:
\begin{align}\label{eq_sol_phi_full}
    \tilde\phi(\tilde r) &=\left\{ I_1\left(\frac{\tilde r}{\delta}\right)\left( A^0_\phi - \int_{\tilde r/\delta}^{+\infty}{K_1(u)(A_\phi u^2 + su + B_\phi)\:
        \diff u} \right) \right. \nonumber\\
        &+\left. K_1\left(\frac{\tilde r}{\delta}\right) \left( B^0_\phi 
        - \int_0^{\tilde r/\delta}{I_1(u)(A_\phi u^2 + su + B_\phi)\: \diff u} \right) \right\} \,,
\end{align}
where $\delta=\sqrt{(\eta+\kappa)/\eta}$.
The velocity then reads:
\begin{align}
    \tilde{v}_r(\tilde r) &= \left(\frac{\tilde{\psi}(\tilde{r})}{\tilde r} + \frac{\diff \phi(\tilde{r})}{\diff \tilde{r}}\right) \cos \theta \label{eq_vr_phi}\\
    \tilde{v}_\theta(\tilde r) &= -\left(\frac{\diff \tilde\psi(\tilde{r})}{\diff \tilde{r}} + \frac{\tilde{\phi}(\tilde{r})}{\tilde r}\right) \sin \theta\ ,
    \label{eq_vt_phi}
\end{align}
where $\tilde\psi$ is given by eq.~\eqref{eq_sol_psi_full} as in~\ref{app_velocity_wo}.
The force-balance condition~\eqref{eq_FB_vector} then imposes:
\begin{align}
    B &= \frac{B_\phi}{\delta} \label{B_phi} \\
    A &+ \frac{A_\phi}{\delta} = \tilde{v}_0\ . \label{A_phi}
\end{align}
Imposing that the divergence of the velocity field---or equivalently the net division rate $k$---does not diverge at the core nor at infinity, we find that $A^0_\phi$, $A_0$, $B^0_\phi$, $B_0$, $B$, and $B_\phi$ must all vanish.
At the center of the defect $\tilde r =0$, the components of velocity then read $\tilde{v}_r(\tilde r = 0) = -(\tilde{v}_0 + s (\pi/4)(1+1/\delta))$ and $\tilde{v}_\theta(\tilde r = 0) = \tilde{v}_0 + s (\pi/4)(1+1/\delta)$. Imposing $\tilde{\mathbf{v}}(\tilde r = 0) = \mathbf{0}$ in the reference frame of the default yields the self-advection velocity given by eq.~\eqref{eq_vdefect_division}.

The divergence of the velocity field is then given by:
\begin{align}
    \nabla \cdot \tilde{\vv} = 
     -\frac{s}{\delta^2}&\left[I_1\left(\frac{\tilde r}{\delta}\right) \int_{\tilde r/\delta}^{+\infty}{K_1(u)\, u \:
        \diff u} \right. \nonumber\\
        &\left.+ K_1\left(\frac{\tilde r}{\delta}\right)\int_0^{\tilde r/\delta}{I_1(u)\,u\: \diff u} -1\right] \cos \theta \ .
        \label{eq_divv_p12}
\end{align}
It is represented in fig.~\ref{fig_div_rate}.

\section{-1/2 defects}\label{app_m12}
\subsection{Velocity field}\label{app_m12_velocity}
This appendix is dedicated to the computation of the velocity field around a  $-1/2$ defect, plotted on fig.~\ref{fig_velocity_field}. We add the superscript `-' to denote the quantities associated to a $-1/2$ defect. Force balance reads:
\begin{equation}
    \eta \Delta \mathbf{v}^- - \mathbf{\nabla} P^- - \xi \mathbf{v}^-  
    -\frac{\zeta \Delta \mu}{2r} \left( -\cos(3\theta)\: \mathbf{e}_r 
    + \sin(3\theta)\: \mathbf{e}_\theta\right)= 0 \ .
    \label{eq_FB_vector_m12}
\end{equation}
Here the defect is immotile, such that the velocity field satisfies:
\begin{align}
    \mathbf{v}^-(+\infty,\theta) &= 0 \label{eq_BCinf_m12}\\
    \mathbf{v}^-(0,\theta) &= 0 \label{eq_BC0_m12} \ .
\end{align}
The curl of eq.~\eqref{eq_FB_vector_m12} gives:
\begin{equation}
    \Delta \left[ \eta \Delta \psi^-  - \xi \psi^-
    \right] = \frac{3}{2}\zeta \Delta \mu \frac{\sin 3\theta}{r^2} \ .
    \label{eq_psi_4th_order_m12}
\end{equation}
Using a normalization by a characteristic length $L=\sqrt{\eta/\xi}$ and a characteristic time $\tau^- = 2\eta/(3|\zeta| \Delta \mu)$, we introduce the dimensionless stream function $\tilde \psi^- = (\tau^-/L^2) \psi^-$ and spatial variable $\tilde r = r/L$. We then have:
\begin{equation}
    \Delta \left[ \Delta \tilde \psi^- - \tilde \psi^-
    \right] = s\frac{\sin 3\theta}{r^2}\ ,
    \label{eq_psi_4th_order_reduced_m12}
\end{equation}
where $s=\textrm{sign}(\zeta)$ and $\psi^-$ is of the form
$\tilde \psi^-(\tilde r,\theta) = \tilde \psi^-(\tilde r) \sin 3 \theta$.
Integrating one Laplace operator in this equation leads to:
\begin{equation}
    \frac{\diff^2 \tilde \psi^-(\tilde r)}{\diff\tilde r^2} +
    \frac{1}{\tilde r}\frac{\diff \tilde \psi^-(\tilde r)}{\diff \tilde r} -
    \left(1+\frac{3}{\tilde r^2}\right)\tilde \psi^-(\tilde r) = A^- \tilde r + \frac{B^-}{\tilde r} -
    s \ .
    \label{eq_psi_dev_m12}
\end{equation}
Solving this equation leads to the following dimensionless velocity field:
\begin{align}
    \tilde v_r^- = s&\left\{ I_3(\tilde r)  
        \int_{\tilde r}^{+\infty}{ K_3(u) u \:
            \mathrm{d}u}\right. \nonumber \\
    &+\left. K_3(\tilde r) 
    \int_0^{\tilde r}{ I_3(u)u\: \mathrm{d}u}
\right\} \cos 3\theta \label{eq_vr_m12} \\
    \tilde v_\theta^- = s&\left\{ \left(
    \frac{3}{\tilde r}I_3(\tilde r)-I_2(\tilde r) \right)  
        \int_{\tilde r}^{+\infty}{ K_3(u) u \:
            \mathrm{d}u}\right. \nonumber \\
    &+\left. \left(
    \frac{3}{\tilde r}K_3(\tilde r)+K_2(\tilde r) \right)
    \int_0^{\tilde r}{ I_3(u)u\: \mathrm{d}u}
\right\} \sin 3\theta \ ,\label{eq8vt_m12}
\end{align}
accounting for the boundary conditions eqs.~\eqref{eq_BCinf_m12} and \eqref{eq_BC0_m12}. This velocity field is plotted on fig.~\ref{fig_velocity_field}b,d.

The pressure field $P^-$ is obtained by taking the divergence of eq.~\eqref{eq_FB_vector_m12}:
\begin{equation}\label{eq_app_Pm_diff}
    \Delta P^- = \frac{3 \zeta \Delta \mu}{2r^2}\cos 3\theta\ ,
\end{equation}
which yields, with the boundary conditions~\eqref{eq_BCinf_m12}--\eqref{eq_BC0_m12}:
\begin{equation}\label{eq_app_Pm}
    P^-(r,\theta) = -\frac{3 \zeta \Delta \mu}{2}\cos 3\theta \ .
\end{equation}

\subsection{Cell division/extrusion}\label{app_m12_division}
We derive in this section the divergence of the velocity field represented on fig.~\ref{fig_div_rate}.
The derivation resembles that for +1/2 defects as presented in section~\ref{sec_division} with a pressure-dependent division rate given by Eq.~\eqref{eq_division_rate_pressure}.
The velocity field is decomposed into $\mathbf{v}^- = \mathbf{\nabla} \times (\psi^-\ \ve_z)+\mathbf{\nabla} \phi^-$. The divergence-free and curl-free parts of the velocity respectively satisfy:
\begin{align}
\Delta \left[\Delta \tilde{\psi}^-  - \tilde{\psi}^-\right] &= s\frac{\sin 3\theta}{\tilde{r}^2}\label{eq_FB_psi_m12} \\
\Delta \left[\Delta \tilde{\phi}^- - \frac{1}{\delta^2}\tilde{\phi}^- \right] &= s\frac{\cos{3\theta}}{\delta^2 \tilde{r}^2} \label{eq_FB_phi_m12}\ .
\end{align}
Solving eq.~\eqref{eq_FB_phi_m12}, we get the divergence profile of the velocity field represented on fig.~\ref{fig_div_rate}b,d as:
\begin{align}
    \nabla \cdot \tilde{\vv}^- = 
     \frac{s}{\delta^2}&\left[I_3\left(\frac{\tilde r}{\delta}\right) \int_{\tilde r/\delta}^{+\infty}{K_3(u)\, u \:
        \diff u} \right. \nonumber\\
        &\left.+ K_3\left(\frac{\tilde r}{\delta}\right)\int_0^{\tilde r/\delta}{I_3(u)\,u\: \diff u}-1\right] \cos 3\theta \ .
        \label{eq_divv_m12}
\end{align}
\end{document}